\documentclass{article}

\title{A Lax pair of the discrete Euler top in terms of quaternions}
\author{Kinji Kimura\\
Graduate School of Informatics, Kyoto University, Kyoto, Kyoto, JAPAN}
\date{}

\begin{document}

\maketitle

\abstract{We proposed the discrete Euler top in 2000. In that paper, exact solutions and conserved quantities are described. However, a Lax pair of our proposed discrete Euler top is not contained. Moreover, the Lax pair is still unknown. In this paper, from a generalized eigenvalue problem, we obtain the Lax pair of the discrete Euler top. In addition, we introduce another Lax pair of the discrete Euler top in terms of quaternions.}

\section{Introduction}
We proposed the discrete Euler top in \cite{HK}. In that paper, exact solutions and conserved quantities are described. However, a Lax pair of our proposed discrete Euler top is not contained. Moreover, the Lax pair is still unknown\cite{Hone}. In this paper, from a generalized eigenvalue problem, we obtain the Lax pair of the discrete Euler top. In addition, we introduce another Lax pair of the discrete Euler top in terms of quaternions.

\section{A generalized eigenvalue problem}
We consider a generalized eigenvalue problem\cite{Z}.
Let $A$ and $B$ as $m \times m$ matrices. 
Let us denote eigenvalues by $\lambda$. 
Define $\phi(\lambda)$ as eigenvectors.
\begin{eqnarray}
A \phi(\lambda) = \lambda B \phi(\lambda), \label{z1}
\end{eqnarray}
such that, 
\begin{eqnarray}
\tilde A T^{(1)}=T^{(2)}A, \tilde B T^{(1)}=T^{(2)}B, \label{z2}
\end{eqnarray}
where $\tilde A$, $\tilde B$, $T^{(1)}$ and $T^{(2)}$ are also $m \times m$ matrices.
We define the eigenvectors $\tilde \phi (\lambda)$ as follows, 
\begin{eqnarray}
\tilde \phi (\lambda) = T^{(1)} \phi (\lambda). \label{z3}
\end{eqnarray}
From the eqs. (\ref{z1}), (\ref{z2}) and (\ref{z3}), we obtain,
\begin{eqnarray}
\tilde A \tilde \phi (\lambda)=\lambda \tilde B \tilde \phi (\lambda).
\end{eqnarray}
From the repetition of the above operation, we can get,
\begin{eqnarray}
& A_n \phi_n (\lambda)=\lambda B_n \phi_n (\lambda), \quad n=0,1,\cdots, \label{r1} \\
& A_{n+1} T^{(1)}_n=T^{(2)}_n A_n, B_{n+1} T^{(1)}_n=T^{(2)}_n B_n, \\
& \phi_{n+1} (\lambda) = T^{(1)}_n \phi_n (\lambda),
\end{eqnarray}
where $A_n$ and $B_n$ are $m \times m$ matrices and $\phi_n (\lambda)$ is 
eigenvectors. 
If we choose $T^{(1)}=B, T^{(2)}=\tilde B$, 
then $\tilde A B= \tilde B A$. 
In this case, $\tilde \phi (\lambda) = B \phi (\lambda)$. 
Moreover, $A_n$ and $B_n$ satisfy,
\begin{eqnarray}
A_{n+1} B_n= B_{n+1} A_n, \label{z4}
\end{eqnarray}
which is a well-known result(see \cite{Suris1,Suris2}).
From the eq. (\ref{r1}), $\lambda$ is the conserved quantity. Then, computing the characteristic polynomial,
\begin{eqnarray}
p(\lambda)=\mbox{det}(\lambda B_n-A_n)
=c_m \lambda^m+\cdots+c_0,
\end{eqnarray}
we can obtain the following conserved quantities,
\begin{eqnarray}
H_0=\frac{c_0}{c_m},\cdots,H_{m-1}=\frac{c_{m-1}}{c_m}.
\label{conserved}
\end{eqnarray}
Therefore, we can regard $A_n$ and $B_n$ as the discrete version of the Lax pair. However, the number of dependent variables of $n$ in $A_n$ and $B_n$ is equal to $2 m \times m$. 
If $A_n$ and $B_n$ are arbitrary matrices, 
we cannot compute dependent variables of $A_n$ and $B_n$ uniquely from the eq. (\ref{z4}). In the next section, as examples of $A_n$ and $B_n$, 
we propose special matrices which consist of only $m \times (m-1)$ dependent variables.

\section{Examples of $A_n$ and $B_n$}

Let $\alpha_{1,1}, \cdots, \alpha_{1,m}, \alpha_{2,2}, \cdots, \alpha_{2,m},
\cdots, \alpha_{m,m},\beta_{1,1}, \cdots, \beta_{1,m}, \beta_{2,2}, \cdots, \beta_{2,m},\cdots, \beta_{m,m}$ be parameters.
Define $x_{1,2}^n, \cdots, x_{1,m}^n$$, x_{2,1}^n,x_{2,3}^n, \cdots, x_{2,m}^n$,$x_{3,1}^n, x_{3,2}^n,x_{3,4}^n,\cdots,x_{3,m}^n,\cdots,$\\
$x_{m,1}^n,\cdots,x_{m,m}^n$ as $m \times (m-1)$ dependent variables of $n$. If we define,
\begin{eqnarray}
&& A_n=\left(
\begin{array}{ccccc}
\alpha_{1,1}  & \alpha_{1,2} x_{1,2}^{n} & \alpha_{1,3} x_{1,3}^{n} & \cdots & \alpha_{1,m} x_{1,m}^{n} \\
\alpha_{1,2} x_{2,1}^{n} & \alpha_{2,2}         & \alpha_{2,3} x_{2,3}^{n} & \cdots & \alpha_{2,m} x_{2,m}^{n} \\
\alpha_{1,3} x_{3,1}^{n} & \alpha_{2,3} x_{3,2}^{n} & \ddots      & \cdots & \alpha_{3,m} x_{3,m}^{n} \\
\vdots      & \vdots      & \ddots      & \ddots & \vdots      \\
\alpha_{1,m} x_{m,1}^{n} & \alpha_{2,m} x_{m,2}^{n} & \cdots      & \alpha_{m-1,m} x_{m,m-1}^{n} & \alpha_{m,m}
\end{array}
\right), \nonumber \\
&& \label{g1} \\
&& B_n=\left(
\begin{array}{ccccc}
\beta_{1,1}  & \beta_{1,2} x_{1,2}^{n} & \beta_{1,3} x_{1,3}^{n} & \cdots & \beta_{1,m} x_{1,m}^{n} \\
\beta_{1,2} x_{2,1}^{n} & \beta_{2,2}         & \beta_{2,3} x_{2,3}^{n} & \cdots & \beta_{2,m} x_{2,m}^{n} \\
\beta_{1,3} x_{3,1}^{n} & \beta_{2,3} x_{3,2}^{n} & \ddots      & \cdots & \beta_{3,m} x_{3,m}^{n} \\
\vdots      & \vdots      & \ddots      & \ddots & \vdots      \\
\beta_{1,m} x_{m,1}^{n} & \beta_{2,m} x_{m,2}^{n} & \cdots      & \beta_{m-1,m} x_{m,m-1}^{n} & \beta_{m,m}
\end{array}
\right) \nonumber, \\
&& \label{g2}
\end{eqnarray}
then we can obtain many kinds of the discrete systems from the eq. (\ref{z4}).

\subsection{Conjecture}

Let $c_{1}, \cdots, c_{m}$,$d_{1}, \cdots, d_{m}$,$f_{1}, \cdots, f_{m}$,$g_{1}, \cdots, g_{m}$ be parameters. \\
Define $x_{1,2}^n, \cdots, x_{1,m}^n$, 
$x_{2,1}^n,x_{2,3}^n, \cdots, x_{2,m}^n$,$x_{3,1}^n, x_{3,2}^n,x_{3,4}^n,\cdots,x_{3,m}^n$,$\cdots$, \\
$x_{m,1}^n,\cdots,x_{m,m}^n$ as $m \times (m-1)$ dependent variables of $n$. As a conjecture, if we define,
\begin{eqnarray}
&& C=\mbox{diag}(c_1,\cdots,c_m), \label{g3} \\
&& D=\mbox{diag}(d_1,\cdots,d_m), \\
&& F=\mbox{diag}(f_1,\cdots,f_m), \\
&& G=\mbox{diag}(g_1,\cdots,g_m), \\
&& X_n=\left(
\begin{array}{ccccc}
0  & x_{1,2}^{n} & x_{1,3}^{n} & \cdots & x_{1,m}^{n} \\
x_{2,1}^{n} & 0     & x_{2,3}^{n} & \cdots & x_{2,m}^{n} \\
x_{3,1}^{n} & x_{3,2}^{n} & \ddots      & \cdots & x_{3,m}^{n} \\
\vdots      & \vdots      & \ddots      & \ddots & \vdots      \\
x_{m,1}^{n} & x_{m,2}^{n} & \cdots      & x_{m,m-1}^{n} & 0
\end{array}
\right), \\
&& A_n=C+F X_n-X_n F, \\
&& B_n=D+G X_n-X_n G, \label{g4}
\end{eqnarray}
where $C,D,F$ and $G$ are diagonal matrices, then we can obtain many kinds of the discrete integrable systems from the eq. (\ref{z4}). 

\section{A Lax pair of the discrete Euler top}

Let $\beta_1,\beta_2,\beta_3$ be parameters.
Define $\omega_{1}^n, \omega_{2}^n, \omega_{3}^n$ as dependent variables of $n$. If we define,
\begin{eqnarray}
&& A_n=\left(
\begin{array}{cccc}
 0            &  \omega_1^{n} &  \omega_2^{n} &  \omega_3^{n} \\
-\omega_1^{n} &             0 &  \omega_3^{n} & -\omega_2^{n} \\
-\omega_2^{n} & -\omega_3^{n} &             0 &  \omega_1^{n} \\
-\omega_3^{n} &  \omega_2^{n} & -\omega_1^{n} &             0 
\end{array}
\right), \label{Euler1} \\
&& B_n=\left(
\begin{array}{cccc}
 1            &  \beta_1 \omega_1^{n} &  \beta_2 \omega_2^{n} &  \beta_3 \omega_3^{n} \\
-\beta_1 \omega_1^{n} &             1 & \beta_3 \omega_3^{n} & -\beta_2 \omega_2^{n} \\
-\beta_2 \omega_2^{n} & -\beta_3 \omega_3^{n} &             1 &  \beta_1 \omega_1^{n} \\
-\beta_3 \omega_3^{n} &  \beta_2 \omega_2^{n} &  -\beta_1 \omega_1^{n} &             1 
\end{array}
\right), \label{Euler2}
\end{eqnarray}
then we can obtain the discrete Euler top,
\begin{eqnarray}
\omega_1^{n+1}-\omega_1^{n}=(\beta_3 -\beta_2 )(\omega_2^{n+1}\omega_3^{n}+\omega_2^{n}\omega_3^{n+1}), \\
\omega_2^{n+1}-\omega_2^{n}=(\beta_1 -\beta_3 )(\omega_3^{n+1}\omega_1^{n}+\omega_3^{n}\omega_1^{n+1}), \\
\omega_3^{n+1}-\omega_3^{n}=(\beta_2 -\beta_1 )(\omega_1^{n+1}\omega_2^{n}+\omega_1^{n}\omega_2^{n+1}),
\end{eqnarray}
which is proposed in \cite{HK} from the eq. (\ref{z4}). 
$A_n$ and $B_n$ in the eqs. (\ref{Euler1}) and (\ref{Euler2}) are obtained as the special case of $A_n$ and $B_n$ in the eqs. (\ref{g1}) and (\ref{g2}).
From $H_0,\cdots,H_{3}$ in the eq. (\ref{conserved}), we can get two conserved quantities of the discrete Euler top.
We can check that $H_0$ and $H_{3}$ are equivalent to the conserved quantities described in \cite{HK}.

\section{Another Lax pair of the discrete Euler top in terms of quaternions}

In addition, we propose another Lax pair of the discrete Euler top in terms of quaternions.
Let $\beta_1,\beta_2,\beta_3$ be parameters.
Define $x_{1}^n, x_{2}^n, x_{3}^n$ as dependent variables of $n$. If we define,
\begin{eqnarray}
&& A^n=\left(
\begin{array}{cccc}
 0            &  x_1^{n} &  x_2^{n} &  x_3^{n} \\
-x_1^{n} &             0 &  -x_3^{n} & x_2^{n} \\
-x_2^{n} &   x_3^{n} &             0 &  -x_1^{n} \\
-x_3^{n} &  -x_2^{n} & x_1^{n} &             0 
\end{array}
\right), \\
&& B^n=\left(
\begin{array}{cccc}
 1            &  \beta_1 x_1^{n} &  \beta_2 x_2^{n} &  \beta_3 x_3^{n} \\
-\beta_1 x_1^{n} &             1 & -\beta_3 x_3^{n} & \beta_2 x_2^{n} \\
-\beta_2 x_2^{n} & \beta_3 x_3^{n} &             1 &  -\beta_1 x_1^{n} \\
-\beta_3 x_3^{n} &  -\beta_2 x_2^{n} &  \beta_1 x_1^{n} &             1 
\end{array}
\right), 
\end{eqnarray}
then we can obtain,
\begin{eqnarray}
x_1^{n+1}-x_1^{n}=(\beta_2 -\beta_3 )(x_2^{n+1}x_3^{n}+x_2^{n}x_3^{n+1}), \\
x_2^{n+1}-x_2^{n}=(\beta_3 -\beta_1 )(x_3^{n+1}x_1^{n}+x_3^{n}x_1^{n+1}), \\
x_3^{n+1}-x_3^{n}=(\beta_1 -\beta_2 )(x_1^{n+1}x_2^{n}+x_1^{n}x_2^{n+1}),
\end{eqnarray}
from the eq. (\ref{z4}).
If we restrict $\beta_1,\beta_2,\beta_3$ and $x_{1}^0, x_{2}^0, x_{3}^0$ to real numbers, then we can rewrite $A^n$ and $B^n$ as follows,
\begin{eqnarray*}
&& A^n=\left((x_1^{n}) {\it \bf i} +  (x_2^{n}) {\it \bf j} + (x_3^{n}) {\it \bf k}\right), \\
&& B^n=\left(1 + (\beta_1 x_1^{n}) {\it \bf i} + (\beta_2 x_2^{n}) {\it \bf j} + (\beta_3 x_3^{n}) {\it \bf k}\right),
\end{eqnarray*}
where ${\it \bf i}$, ${\it \bf j}$ and ${\it \bf k}$ are the fundamental quaternion units.

\section{Conclusion}

From the generalized eigenvalue problem, we obtained the Lax pair of the discrete Euler top proposed in \cite{HK}. In addition, we introduced another Lax pair of the discrete Euler top in terms of quaternions.

\end{document}